\documentclass{aa}  

\usepackage[backref,breaklinks,colorlinks,linkcolor=blue,urlcolor=blue,citecolor=blue]{hyperref}
\usepackage[varg]{txfonts}

\usepackage{graphicx}
\usepackage{bm}
\newcommand{\msun}{\,\mathrm{M}_{\sun}}
\newcommand{\mpc}{\,\mathrm{Mpc}\,h^{-1}}

\begin{document}

\title{Extragalactic archaeology with the C, N, and O chemical abundances}

\author {Fiorenzo Vincenzo \thanks {f.vincenzo@herts.ac.uk} \and Chiaki Kobayashi\thanks {c.kobayashi@herts.ac.uk}} \institute{
  Centre for Astrophysics Research, University of Hertfordshire, College Lane, Hatfield, AL10 9AB, UK }

\date{Received 1 December 2017 / Accepted 7 February 2018}

\abstract{We predict how the C, N, and O abundances within the interstellar medium of galaxies evolve as functions of the galaxy star formation history (SFH). 
We adopt a hydrodynamical cosmological simulation, focusing on three star-forming disc galaxies with different SFHs. 
By assuming failed supernovae, we can 
predict an increasing trend of the gas-phase N/O--O/H abundance diagram, which was not produced in our previous 
simulations without failed supernovae. At high redshifts, contrary to the predictions of classical chemical evolution models with instantaneous mixing approximation, 
 we find almost flat trends in the N/O--O/H diagram, which are  
due to the contribution of intermediate-mass stars together with an inhomogeneous chemical enrichment. 
Finally, we also predict that the average N/O and C/O steadily increase as functions of time, while the average C/N decreases, due to the mass and metallicity dependence of the yields of asymptotic giant branch stars; such variations are more marked during more intense star formation episodes. Our predictions on the CNO abundance evolution can be used to study the SFH of disc galaxies with the James Webb Space Telescope.
} 


\titlerunning{Extragalactic archaeology with the C, N and O abundances}
\authorrunning{F. Vincenzo and C. Kobayashi}

\maketitle

\section{Introduction} \label{sec:intro}

High-resolution multi-object spectrographs have made it possible 
to retrieve accurate information about the kinematics, chemical composition, ages and other fundamental 
physical properties for a very large ensemble of stars, putting strong constraints on the formation and evolution of our Galaxy 
(e.g. \citealt{bensby2014,hayden2015,lindegren2016}). 
Nevertheless, as soon as we enter into the broad field of extragalactic astronomy, our comprehension 
about how galaxies have formed and evolved with time becomes more uncertain, mostly because their 
global observed physical properties have usually been inferred from their integrated stellar light or emission lines. 

Following the pioneering works by \citet{garnett1990} and \citet{vilacostas1993}, 
emission line diagnostics are capable of measuring C, N, and O abundances, 
and are being applied to extragalactic systems, providing also the spatial distribution of the elemental abundances 
(e.g. \citealt{pilyugin2010,sanchez-menguiano2016,belfiore2017,toribio2017,amorin2017}). 
This {blossoming} of observational data will open a new window of interest for galactic astroarchaeology studies. 
 In this respect, chemical evolution models -- when included within cosmological hydrodynamical simulations and combined with 
spectrophotometric stellar population synthesis codes -- can uniquely provide a powerful tool to unveil the past evolutionary 
history of galaxies. Only by dealing with the abundances of specific chemical elements (and hence only by shelving the generic 
observable ``metallicity''), can we define sensitive chemical diagnostics of the entire star formation history (SFH) of the galaxy.

In this Letter, we present the results of our study on a sample of star-forming disc galaxies within 
an up-to-date cosmological hydrodynamical simulation, including a detailed routine for the galaxy chemical evolution \citep{kobayashi2007}.  
We focus our analysis on the evolution of three examples of disc galaxies, representative of the different stellar systems within our catalogue, and 
show how their gas-phase chemical abundances of C, N, and O 
evolve as functions of the galaxy SFH.  
In Section \ref{sec:model}, we briefly summarise the main characteristics 
and the setup of our simulation. 
In Section \ref{sec:results}, we present the inhomogeneous enrichment and the global gas-phase CNO abundance evolution of our disc galaxies. 
Finally, in Section \ref{sec:conclusions}, we draw our conclusions.

\begin{table*}[t!]
\renewcommand{\thetable}{\arabic{table}}
\centering
\scalebox{0.7}{
\begin{tabular}{cllcccccrccc}
\hline
\hline
Galaxy name & redshift & $M_{\star}$  & $f_{\text{gas}}$ &  $\tau_{\star,1/2} $ & $\langle \tau_\star \rangle_{V}$ & $\langle Z_{\star} \rangle_{V}$ & 
   $r_{h,V} $ & $\log\big(\langle\text{N/O}\rangle_{V}\big) $ & $ \log\big(\langle\text{C/O}\rangle_{V}\big)$ & $ \log\big(\langle\text{C/N}\rangle_{V}\big)$         \\
                      &             &  $[\text{M}_{\sun}]$  &                         &  $[\text{Gyr}]$  &  $[\text{Gyr}]$  &                   &
  $[\text{kpc}]$ & $\text{dex}$ &  $\text{dex}$ &  $\text{dex}$    \\                           
\hline
Galaxy A  & $z=0$     & $3.62\times10^{10}$     &  $0.35$  & $6.69$    & $3.27$   &  $0.012$   & $4.58$    & $-0.85$ & $-0.45$ & $0.40$    \\
                & $z=0.5$  & $2.73\times10^{10}$     &  $0.43$  & $3.28$    & $1.41$   &     $0.012$   & $2.63$ & $-0.84$ & $-0.43$  & $0.41$       \\
                & $z=1$     & $1.77\times10^{10}$     &  $0.35$  & $1.83$    & $0.80$   &    $0.011$   & $1.63$  & $-0.91$ & $-0.46$  & $0.45$      \\
                & $z=2$     & $6.72\times10^{9}$     &  $0.53$  & $0.82$    & $0.42$   &    $0.008$   & $1.02$ & $-1.00$ & $-0.53$  & $0.46$       \\   
                & $z=3$     & $2.64\times10^{9}$     &  $0.63$  & $0.45$    & $0.22$   &    $0.006$   & $0.67$ & $-1.12$ & $-0.69$  & $0.43$       \\   
                & $z=4$     & $7.20\times10^{8}$     &  $0.69$  & $0.13$    & $0.06$   &    $0.003$   & $0.27$   & $-1.39$ & $-0.91$ & $0.48$      \\   
                \hline
Galaxy B &  $z=0$    & $1.60\times10^{10}$     & $0.37$   &  $5.53$  & $2.32$    &   $0.009$  & $3.52$   & $-0.87$ & $-0.46$  & $0.41$     \\
                & $z=0.5$  & $1.13\times10^{10}$     & $0.34$    &  $1.97$  & $1.17$    &  $0.013$  & $1.40$  & $-0.87$ & $-0.43$   & $0.44$     \\
                & $z=1$     & $4.80\times10^{9}$     & $0.57$    &  $1.30$  & $0.55$    &   $0.008$  & $1.31$ & $-1.05$ & $-0.55$  & $0.50$        \\
                & $z=2$     & $5.18\times10^{8}$     & $0.87$    &  $0.89$  & $0.07$    &   $0.002$  & $0.52$   & $-1.43$ & $-0.83$  & $0.59$     \\
                 \hline
Galaxy C & $z=0$     & $9.93\times10^{9}$    & $0.48$  &   $4.63$   & $1.99$    &      $0.010$  & $4.63$ & $-0.90$ & $-0.45$ & $0.45$     \\
                & $z=0.5$  & $5.32\times10^{9}$    & $0.57$  &   $1.66$   & $0.75$    &     $0.008$  & $1.66$  & $-1.05$ & $-0.55$  & $0.51$   \\
                & $z=1$     & $2.08\times10^{9}$    & $0.76$  &   $1.16$   & $0.52$    &     $0.005$  & $1.17$  & $-1.16$ & $-0.61$  & $0.54$   \\
\hline

\end{tabular}
}
\caption{The main properties of our reference galaxies as functions of redshift. Columns report the 
following quantities: i) galaxy name; ii) redshift; iii) total galaxy stellar mass at the present time; iv) fraction of gas within the galaxy with respect to the total 
baryonic mass (stellar plus gas content); v) age at which half of the galaxy stellar mass originated; 
vi) average V-band luminosity-weighted age; 
vii) average stellar metallicity; viii) galaxy disc half-light radius; ix-x-xi) average 
stellar N/O, C/O and C/N ratios. } \label{tab1}
\end{table*}

\section{Simulation model and assumptions} \label{sec:model}

 All the results of this Letter are obtained by running and analysing the outcome of a cosmological hydrodynamical simulation 
 based on the \textsc{Gadget-3} code \citep{springel2005}, which uses the smoothed particle 
 hydrodynamics (SPH) method. Our code was developed by \citet{kobayashi2007} to include the relevant physical processes related 
 to the star formation activity and chemical evolution of galaxies (see also \citealt{taylor2014} for the details). 
 
\begin{figure}[t!] 
\centering
\includegraphics[scale=0.45]{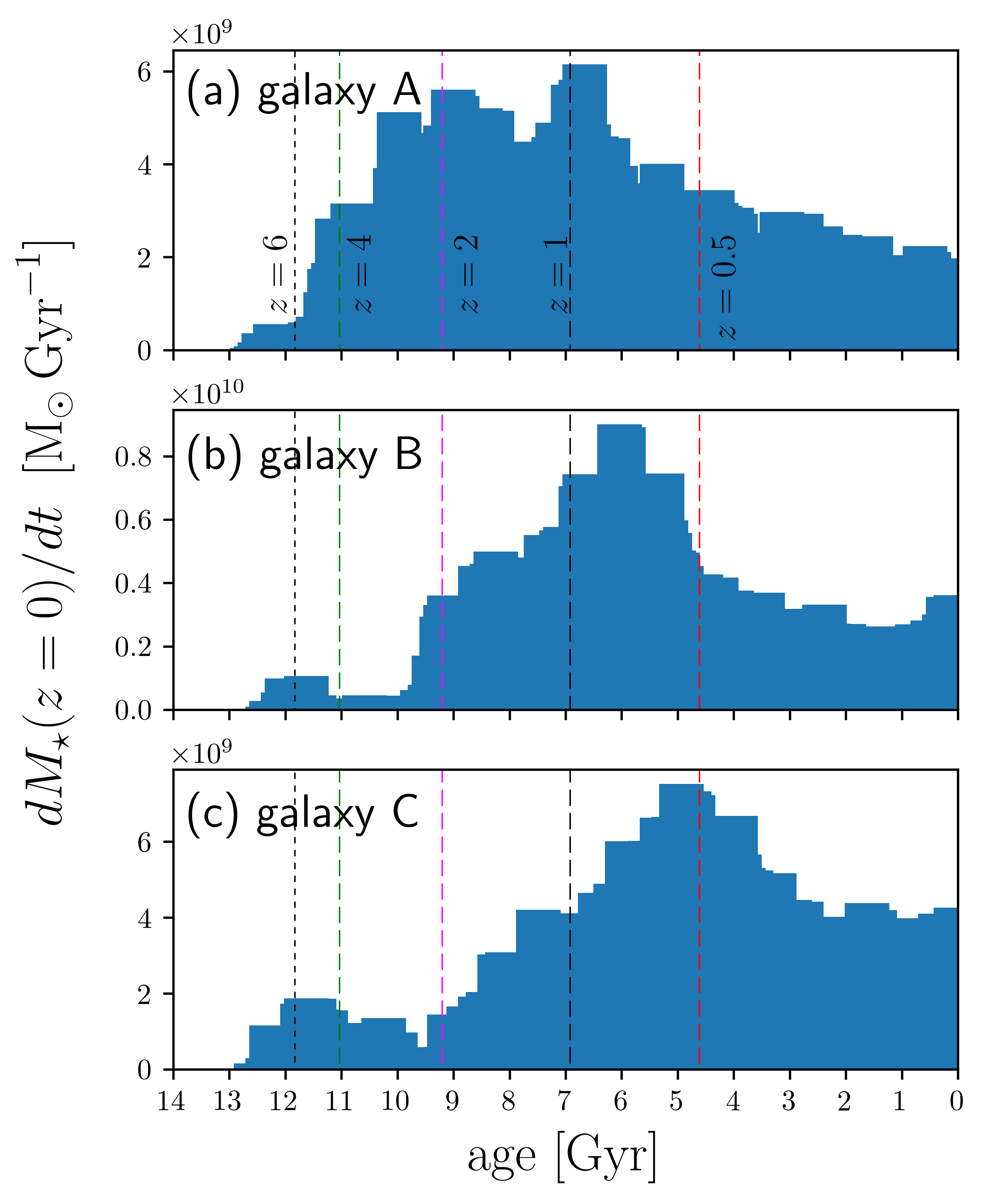} 
\caption{The distribution of 
the present-day total stellar mass as a function of the star particle age. 
Each panel corresponds to one of our reference galaxies.  }
\label{fig:age_distributions}
\end{figure}

\begin{figure}[t!] 
\centering
\includegraphics[scale=0.45]{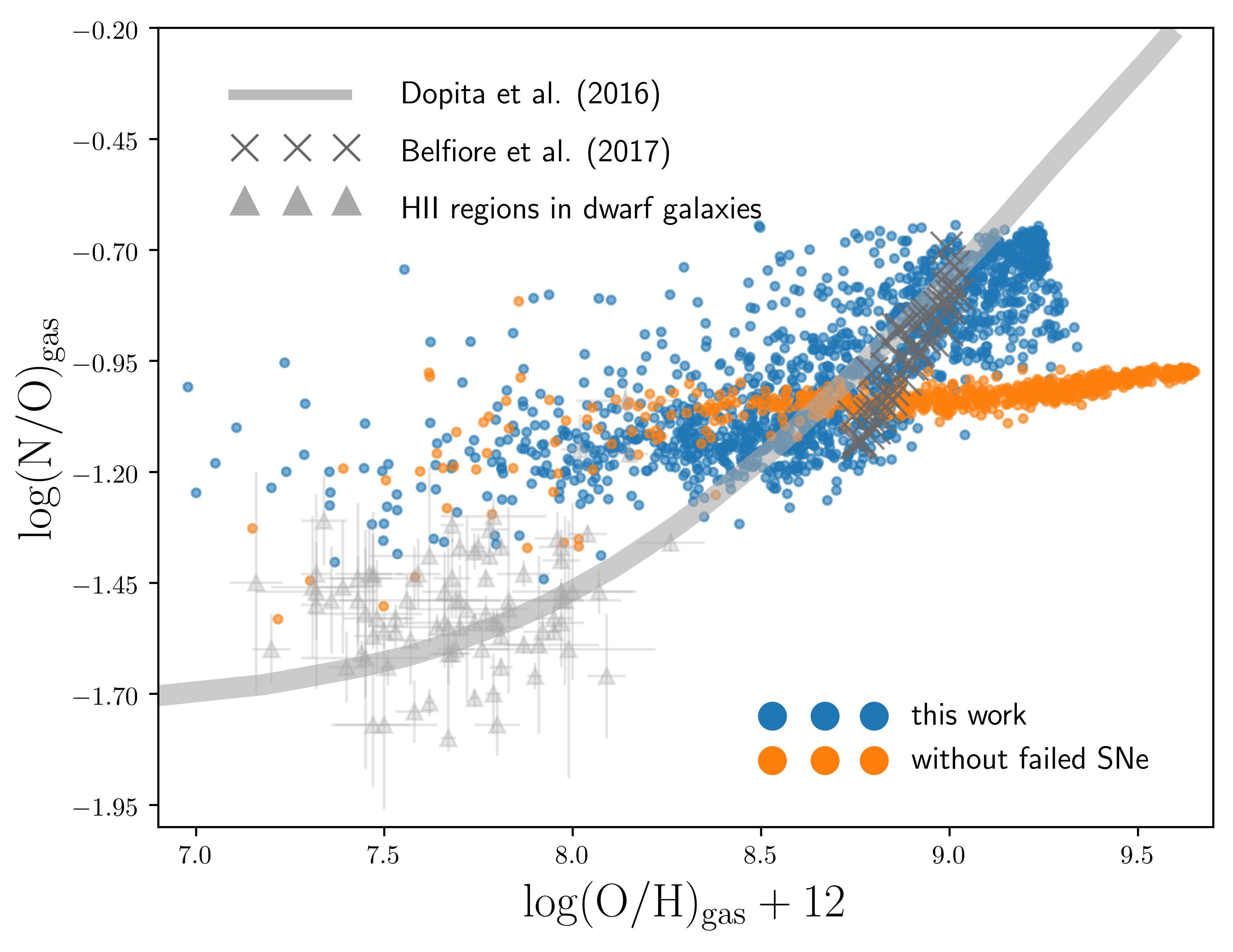} 
\caption{The predicted gas-phase $\log{({\text{N/O}})}$ vs.  $\log{({\text{O/H}})}+12$ 
abundance pattern in Galaxy A with failed SNe (light blue points; this work) and without failed SNe (orange points). 
The solid grey line corresponds to the average relation obtained from various observations by 
\citet[see references therein]{dopita2016}, the grey 
points with error bars to a compilation of data of individual \textsc{Hii} regions in 
blue diffuse dwarf galaxies \citep{berg2012,izotov2012,james2015}, and the dark grey crosses to 
 \citet{belfiore2017} from a sample of spatially resolved galaxies from the MaNGa survey. }
\label{fig:comparison}
\end{figure}

{\it Chemical enrichment} ---
The chemical evolution routine is the same as the one 
 originally developed by \citet{kobayashi2004} for another SPH code, and gives an excellent agreement with the 
 elemental abundance distributions as observed in the Milky Way (MW; \citealt{kobayashi2011a}). 
In our model, we adopt the stellar yield set by \citet{kobayashi2011b} with failed Supernovae (SNe) 
at mass $m \ge 25\,\text{M}_\sun$ and metallicity $Z \ge 0.02$, which are required from the observations 
of nearby SNe (e.g. \citealt{smartt2009}; see also the theoretical study of \citealt{muller2016}).
In failed SNe, all synthesised O and heavier elements fall back into the black hole, except for H, He, C, N and F, 
which are synthesised in the outermost layers of the SN ejecta 
(Kobayashi et al., in prep.). Following \citet{kobayashi2006}, half of stars with $m \ge 25\,\text{M}_\sun$ are assumed 
to explode as hypernovae. For the asymptotic giant branch (AGB) stars, we assume the same yields as in \citet{kobayashi2011b}. 
Finally, we assume the \citet{kroupa2008} initial mass function (IMF), defined in the mass range 
$0.01\le m \le 120\,\text{M}_{\sun}$, and the same stellar lifetimes 
as in \citet{kobayashi2004}.

{\it The photo-chemical code} ---
In order to compute the average luminosity-weighted ages and metallicities of the galaxies in the simulation, as well as their half-light radii, 
we have developed a stellar population synthesis model\footnote{\url{stri-cluster.herts.ac.uk/~fiorenzo}}, which is based 
on the one originally developed by \citet{vincenzo2016b}, assuming the \textsc{PARSEC} stellar evolutionary tracks 
\citep{bressan2012}.

{\it Simulation setup} ---
Our assumed cosmological model  is the standard $\Lambda$-cold dark matter Universe, with the same cosmological 
parameters as given by the nine-year Wilkinson Microwave Anisotropy Probe \citep{hinshaw2013}. 
We follow the evolution of a cubic volume of the Universe, with periodic boundary conditions and 
side $\ell_{\mathrm{U}}=10\mpc$, in comoving units.  
The initial conditions are the same as in \citet{kobayashi2007}, 
with improved resolution; $N_{\mathrm{DM}}=N_{\mathrm{gas}}=128^3$ both 
for dark matter (DM) and gas particles. 
The feedback from the star formation activity and chemical evolution (i.e. the thermal energy and 
the nucleosynthetic products from stellar winds and SN explosions) 
is distributed to a number $N_{\mathrm{FB}}=576$ of neighbour gas particles, weighted by the smoothing kernel.  
These parameters are determined to match the observed cosmic star formation rate (SFR; \citealt{hopkins2006,madau2014}). 
Our simulation has the following mass resolutions: 
$M_{\mathrm{DM}} \approx3.097\times10^{7}\,h^{-1}\msun$ and $M_{\mathrm{gas}}=6.09\times10^{6}\,h^{-1}\msun$, 
for the DM and gas mass components, respectively. Finally, we assume a gravitational 
softening length $\epsilon_{\mathrm{gas}}\approx0.84\,h^{-1}\;\mathrm{kpc}$, in comoving units. 

\begin{figure}[t!]
\centering
\includegraphics[scale=0.5]{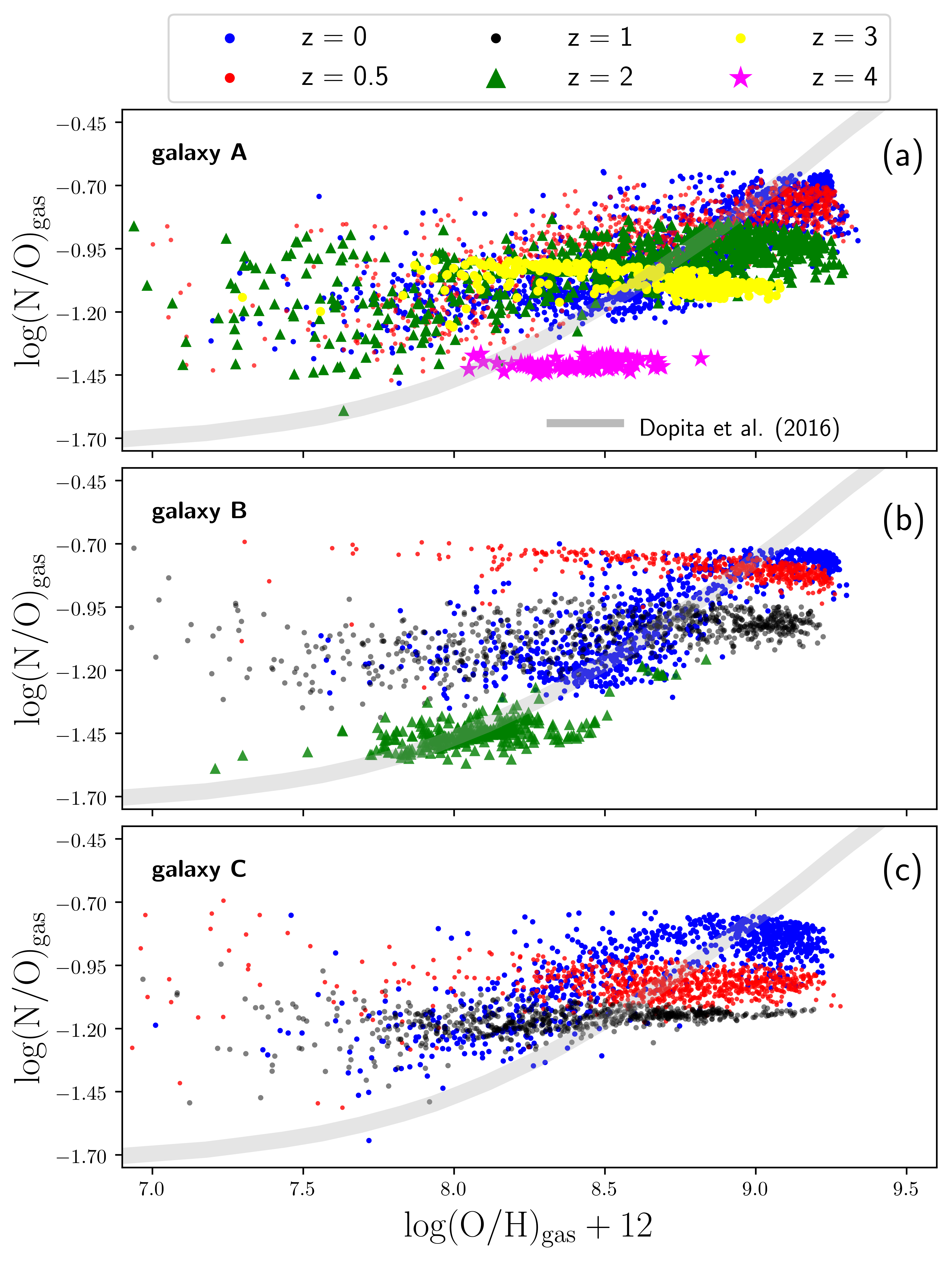} 
\caption{The redshift evolution of the predicted gas-phase $\log{({\text{N/O}})}$ vs.  $\log{({\text{O/H}})}+12$ 
abundance patterns within our three reference galaxies. Galaxy A, B, and C clearly show different chemical evolution of the ISM abundances with redshift.
We compare our simulation with the observed average N/O--O/H relation from \citet[solid grey line]{dopita2016}. }
\label{fig:abundances_redshift}
\end{figure}

\section{Results} \label{sec:results}

We analyse the outcome of our simulation, by creating a catalogue of stellar systems at redshift $z=0$. 
We firstly make use of the friend-of-friends group-finding algorithm with adaptive hierarchical refinement \textsc{Rockstar} 
\citep{behroozi2013} to create a catalogue of DM halos at redshift $z=0$; 
from this catalogue, we isolate and analyse all the embedded stellar systems which 
are spatially clustered and contain a sufficient number of both star and gas particles, 
so as to draw well sampled chemical abundance patterns. 
We exclude all the stellar systems which are undergoing a major merger at the present time. 
In total, we select $\sim35$ galaxies with the aforementioned broad and simple criteria, lying within DM halos 
with virial mass in the range $11.0 \le \log(M_{\mathrm{DM}}/\mathrm{M}_{\sun}) \le 13.0 \;\mathrm{dex}$ (Vincenzo et al., in prep.). 
 
In this Letter, we present the results of our 
analysis for three examples of star-forming disc galaxies, which have 
total stellar masses roughly of the same order of magnitude but are representative of 
distinct characteristic SFHs that are seen in the other disc galaxies in the simulation. 
In Table \ref{tab1}, we summarise the redshift evolution of the main predicted characteristics of our reference galaxies. 
The predicted stellar age distribution function of our galaxies is shown in Figure \ref{fig:age_distributions}. 
When passing from Galaxy C to Galaxy A, the emitted light at the present time is dominated by older stellar populations; 
the chemical enrichment also differs from galaxy to galaxy and 
all our reference galaxies end up at the present time by having almost the same average metallicity 
($\langle Z_{\star} \rangle \sim0.01$), following the observed mass-metallicity relation.

In Figure \ref{fig:comparison}, we show our predictions at redshift $z=0$ for 
the gas-phase $\log{(\text{N/O})}$ versus 
$\log{(\text{O/H})}+12$ abundance pattern in Galaxy A. 
Our results with failed SNe (light blue points) are compared with the same simulation but with the original 
\citet{kobayashi2011b}'s yields without failed SNe (orange points); each simulation 
point in the figure corresponds to one location within the galaxy. 
To understand the abundance patterns, we briefly recall that nitrogen is 
mostly synthesised by intermediate-mass stars 
by hot-bottom burning during the AGB phase. 
Most of the nitrogen from AGB stars is secondary and its stellar yields steadily increase as functions of 
the initial stellar metallicity.  All mass ranges of stars can also produce primary N, converted from $^{12}$C 
by the CNO cycles (see also \citealt{vincenzo2016a} and the references therein). 
Finally, oxygen is mostly produced by core-collapse SNe with very short lifetimes ($\sim10^{6}\,\mathrm{yr}$). 

In Figure \ref{fig:comparison}, with failed SNe, we find an increasing trend (of `secondary' production) at the metal-rich side of the N/O versus O/H diagram, 
predicting a similar slope as in observations. 
At very high metallicity, the increase may be slightly smaller than in the observed relation, which may be because stellar yields 
 from super-solar AGB stars are not available \citep{vincenzo2016a}. We also note that N and O might also be affected by dust depletion, 
 which is not taken into account in our work. If we do not assume failed SNe, we have a much flatter slope at high metallicity, 
 which is inconsistent with observations. 

In Figure \ref{fig:abundances_redshift}, we show how the N/O--O/H relation evolves with redshift in 
our reference galaxies. There is a large scatter particularly at low O/H, 
which is due to inhomogeneous enrichment; in fact, there are gas particles that have 
gathered the nucleosynthetic products of only a few core-collapse SNe or AGB stars in the past 
(see \citealt{kobayashi2011a} for more details). 
When the galaxy is still experiencing {its earliest evolutionary stages ($M_{\star}\sim10^{8}\,\text{M}_{\sun}$)}, the gas-phase 
N/O ratios are almost constant with metallicity, suggesting the presence of a plateau of `primary' production ($z\sim4$ for Galaxy A; 
 $z\sim2$ for Galaxy B). This is not produced in classical one-zone chemical evolution models with instantaneous mixing approximation, 
 which systematically underestimated N/O at very low metallicity and had to assume an artificial primary N production by massive stars 
(e.g. \citealt{matteucci1986,chiappini2005,vincenzo2016a}) to reproduce 
the observations at low metallicity in dwarf galaxies \citep{pilyugin2010,berg2012,izotov2012,james2015}, 
damped-Ly$\alpha$ (DLA) absorption systems 
(e.g., \citealt{pettini2002,pettini2008}), and MW halo stars (e.g., \citealt{spite2005}). 

We note that our predicted N/O plateau resides 
at larger O/H than in the \textsc{Hii} regions of nearby dwarf galaxies, meaning that our reference galaxies were already 
more metal-rich at high redshifts than nearby dwarf galaxies. In the adopted yields, N/O from core-collapse SNe is much lower than in our 
high-redshift galaxies ($\sim - 1.45\,\text{dex}$), but in the simulation 
there is already a significant contribution from AGB stars, which is the difference from the one-zone models. 
At lower redshifts ($M_{\star}\gtrsim10^{9}\,\text{M}_{\sun}$), the plateau exists but the value quickly reaches 
$\log(\text{N/O})\sim-1.2\,\text{dex}$ with a larger contribution from AGB stars. 
Galaxy C is already at this phase when it is formed by a major merger at $z\sim2$.  We note that the plateau value would be even higher with the effects of stellar rotation.

\begin{figure}[t!]
\centering
\includegraphics[scale=0.45]{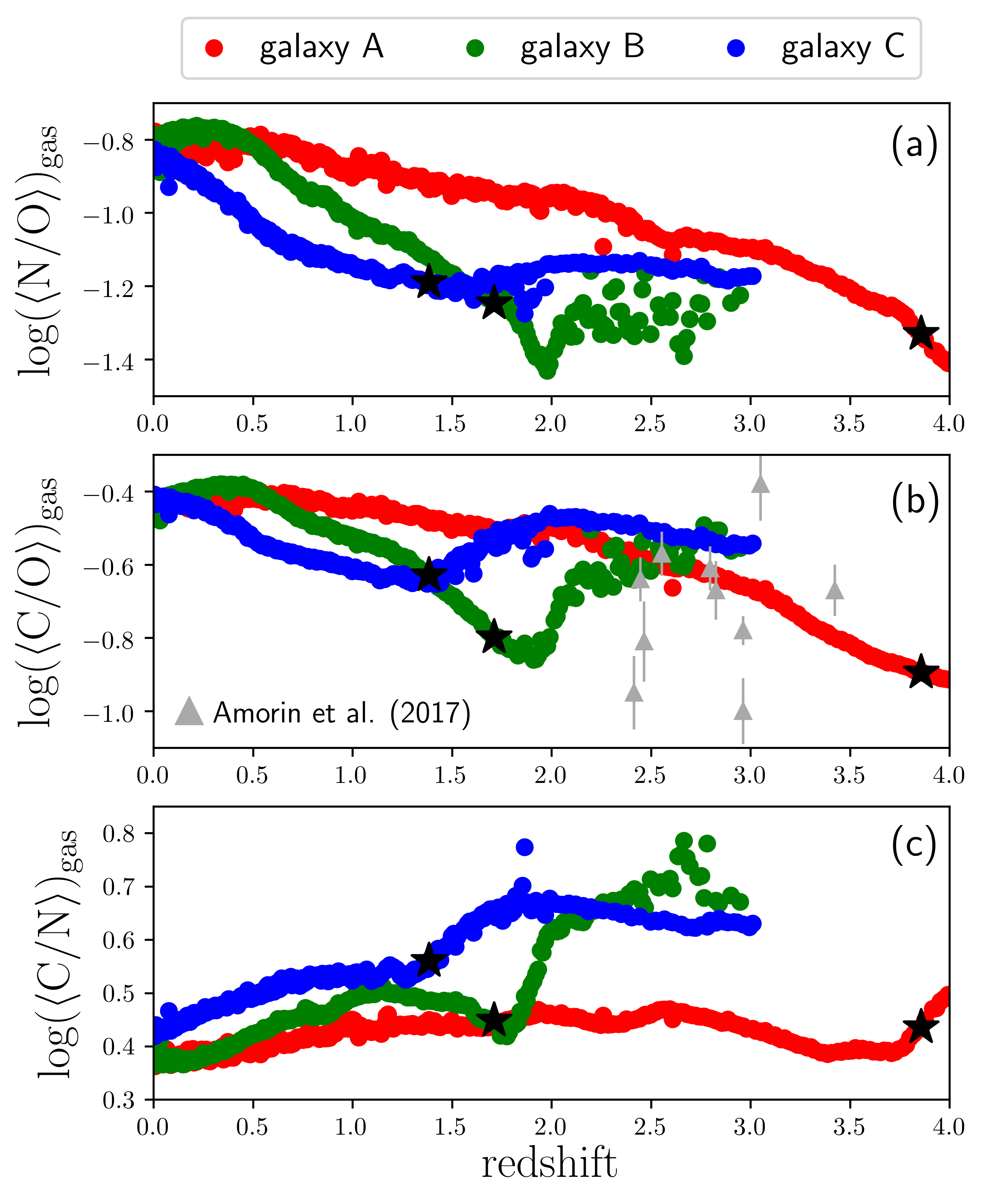} 
\caption{(a-c) The average SFR-weighted gas-phase $\log{({\text{N/O}})}$, $\log{({\text{C/O}})}$ and $\log{({\text{C/N}})}$ as 
functions of redshift. We only show the predictions of our simulation for the redshifts 
when the galaxy stellar mass $M_{\star}(z)\ge10^{8}\,\text{M}_{\sun}$.  The black star symbol on each track 
marks the redshift when $M_{\star} \approx 10^{9}\,\text{M}_{\sun}$. {The grey triangles with the error bars 
correspond to the observational data of \citet{amorin2017}. }
}
\label{fig:average_vs_tau}
\end{figure}

In Figure \ref{fig:average_vs_tau}, we show how the average SFR-weighted gas-phase CNO abundances vary as functions of 
redshift for our reference galaxies; this diagram can be used for constraining the galaxy SFH. We only show the predictions
of our simulation for the evolutionary 
times when the galaxy total stellar mass $M_{\star}(z)\ge 10^{8}\,\text{M}_{\sun}$.  
In Figure \ref{fig:average_vs_tau}(a) the increase of the average N/O ratios turns out to be larger, corresponding to a more intense star formation episode 
(high-$z$ in Galaxy A; low-$z$ in Galaxy C). 
This is the typical imprint of strong chemical enrichment of secondary N from AGB stars.

In Figure \ref{fig:average_vs_tau}(b-c), we show how the average SFR-weighted gas-phase $\log{({\text{C/O}})}$ and 
$\log{({\text{C/N}})}$ evolve with redshift. 
Carbon is mostly synthesised by low-mass stars with masses in the range 
$1\lesssim m \lesssim 3\;\text{M}_{\sun}$ (e.g. see \citealt{kobayashi2011b}). 
As for nitrogen, carbon can also be produced by core-collapse SNe, on short time-scales after the star formation event. 
With the IMF weighting, half of the C producers in galaxies are low-mass stars. 
{The metallicity dependence of the C yields is opposite with respect to the metallicity dependence of the secondary N; 
while the produced secondary N steadily increases as a function of metallicity, the C yield steadily decreases. } 

In Figure \ref{fig:average_vs_tau}(c), at high redshift, in the earliest galaxy evolutionary stages, we mostly witness at the chemical enrichment from massive stars with high C/N. 
The average C/N is subsequently predicted to sharply decrease at the onset of massive AGB stars. 
Then, C/N temporarily increases due to the C production of low-mass AGB stars; 
finally, C/N decreases again because of the opposite metallicy-dependence 
of the C and N stellar yields. {Currently, observations at high redshifts 
are available for C/O ratios \citep{amorin2017}, which roughly overlap with our model curves. 
The very low C/O ratios in \citet{amorin2017} are consistent with the rapid and intense star formation in Galaxy A, but at much lower 
redshifts. }

\section{Conclusions} \label{sec:conclusions}

In this Letter, we have shown how the gas-phase C, N, and O abundances evolve as functions of the galaxy SFH. 
All our results have been obtained by analysing three examples of star-forming disc galaxies within a full cosmological 
hydrodynamical simulation, including up-to-date nucleosynthetic yields and a standard (non-variable) IMF.

Our simulation predicts an almost flat trend of N/O versus O/H at high redshifts, which is caused 
by an inhomogeneous enrichment of the galaxy ISM with a significant contribution of AGB stars 
at low metallicity. Our simulation represents an improvement with respect to previous chemical evolution 
models, both because the N/O--O/H relation was studied with simple one-zone models with instantaneous mixing approximation 
and because we did not assume any 
artificial primary N production by massive stars \citep{matteucci1986,chiappini2005,vincenzo2016a}. 
We note that the N/O plateau would be even higher if rapidly rotating massive stars were included.

{We conclude that failed SNe are a very promising scenario to explain the 
observed trend of N/O versus O/H. The inclusion of super-solar AGB stars might represent a further improvement of our model, 
since they can produce even larger amounts of secondary N as the metallicity increases, helping to 
explain the further increase of N/O at very high O/H}.

The evolution of the gas-phase CNO abundances is driven by the chemical enrichment from AGB stars and 
core-collapse SNe, 
which create an increasing trend of the N/O and C/O 
towards higher metallicities by present. 
The predicted increase of the average N/O  and C/O with redshift 
turns out to be faster within galaxies experiencing a more intense star formation activity during the 
considered redshift interval. Finally, in the earliest
stages of galaxy evolution, we predict the largest 
C/N ratios and the lowest N/O ratios.

It is not easy to obtain both C and N for the same sample of galaxies with current observations, 
since the N/O and C/O ratios can only be estimated for $z\lesssim2.5$ (optical rest frame) and $z\gtrsim2-3$ (UV rest frame), respectively; 
in the near future, however, it will be possible to have measurements of C/N within galaxies at $z\lesssim6$ with the Near-Infrared Spectrograph on the James Webb Space Telescope.

\section*{Acknowledgments}
We thank V. Springel for providing 
\textsc{Gadget-3}, and R. Maiolino, R. Amor\'{i}n, P. Taylor and S. Smartt for stimulating discussions. 
We thank an anonymous referee for his/her comments. 
FV acknowledges funding from the UK 
STFC through grant ST/M000958/1. 
This research has made use of the DiRAC HPC facility in Durham, UK, supported by 
STFC and BIS.

\end{document}